\documentclass[notitlepage,nofootinbib,preprintnumbers,amssymb,superscriptaddress]{revtex4-2}
\usepackage{amsfonts,amssymb,mathtools,graphicx,color,bm,bbm}
\definecolor{ultramarine}{rgb}{0.07, 0.04, 0.56}
\definecolor{cadmiumgreen}{rgb}{0.0, 0.42, 0.24}
\definecolor{indigo(dye)}{rgb}{0.0, 0.25, 0.42}
\usepackage[linktocpage=true]{hyperref}
\hypersetup{
colorlinks=true,
citecolor=ultramarine,
linkcolor=cadmiumgreen,
urlcolor=indigo(dye),
}

\usepackage{autobreak}
\newcommand{\D}{{\rm d}}
\newcommand{\DX}[1]{{\cal X}_{#1}} 
\newcommand{\DXup}[1]{{\cal X}^{#1}}  
\newcommand{\bDX}[1]{\bar{\cal X}_{#1}}

\newcommand{\tri}{{}^{(3)}\!e}
\newcommand{\btri}{{}^{(3)}\!\bar{e}}
\newcommand{\tridot}{{}^{(3)}\!\dot{e}}

\newcommand{\fr}[2]{\frac{#1}{#2}}
\newcommand{\pa}{\partial}
\newcommand{\ti}{\tilde}
\newcommand{\na}{\nabla}
\newcommand{\bra}[1]{\left( #1 \right)}  
\newcommand{\brb}[1]{\left[ #1 \right]}  
  
\newcommand{\be}{\begin{equation}}  
\newcommand{\ee}{\end{equation}}
\newcommand{\bem}{\begin{bmatrix}}
\newcommand{\eem}{\end{bmatrix}}

\newcommand{\ga}{\gamma}

\newcommand{\mn}{{\mu \nu}}

\newcommand{\mF}{\mathcal{F}}

\newcommand{\mZ}{\mathcal{Z}}

\begin{document}

\preprint{YITP-22-134}

\title{Consistency of matter coupling in modified gravity}

\author{Kazufumi Takahashi}
\affiliation{Center for Gravitational Physics and Quantum Information, Yukawa Institute for Theoretical Physics, Kyoto University, 606-8502, Kyoto, Japan}

\author{Rampei Kimura}
\affiliation{Waseda Institute for Advanced Study, Waseda University, 1-6-1 Nishi-Waseda, Shinjuku, Tokyo 169-8050, Japan}

\author{Hayato Motohashi}
\affiliation{Division of Liberal Arts, Kogakuin University, 2665-1 Nakano-machi, Hachioji, Tokyo 192-0015, Japan}

\begin{abstract}
Matter coupling in modified gravity theories is a nontrivial issue when the gravitational Lagrangian possesses a degeneracy structure to avoid the problem of the Ostrogradsky ghost.
Recently, this issue was addressed for bosonic matter fields in the generalized disformal Horndeski class, which is so far the most general class of ghost-free scalar-tensor theories obtained by performing a higher-derivative generalization of invertible disformal transformations on Horndeski theories.
In this paper, we clarify the consistency of fermionic matter coupling in the generalized disformal Horndeski theories.
We develop the transformation law for the tetrad associated with the generalized disformal transformation to see how it affects the fermionic matter coupling.
We find that the consistency of the fermionic matter coupling requires an additional condition on top of the one required for the bosonic case.
As a result, we identify a subclass of the generalized disformal Horndeski class which allows for consistent coupling of ordinary matter fields, including the standard model particles.
\end{abstract}

\maketitle

\section{Introduction}\label{sec:intro}

Modified gravity theories, as the name suggests, refer to modifications/extensions of general relativity.
There are (at least) several motivations to explore modified theories of gravity.
First, general relativity is not an ultraviolet complete theory, and hence one expects that it has to be modified at some high energy scale.
Modified gravity theories could also address mysteries in cosmology, e.g., inflation, dark energy, and dark matter.
Besides these, modifying/extending general relativity leads to a deeper understanding of general relativity itself.
With these motivations, innumerable modified gravity models have been proposed up to now.
From the perspective of testing gravity, such modified gravity models are useful for comparison with general relativity~\cite{Koyama:2015vza,Ferreira:2019xrr,Arai:2022ilw}.
Indeed, effects of the modification would be encoded in, e.g., cosmic microwave background, large-scale structure, black holes, and neutron stars.
These phenomena are a key to distinguish modified gravity models from general relativity by observations.

Being a simplest extension of general relativity, scalar-tensor theories (i.e., those written in terms of the metric and a single scalar field) have been extensively studied to this date.
Despite their simplicity, they are expected to capture essential aspects of generic modified gravity models.
Indeed, many modified gravity models contain some additional scalar degree of freedom (DOF) at least effectively.
Also, scalar-tensor theories offer a rich phenomenology in cosmology and compact objects.
In order to treat various scalar-tensor theories in a unified manner, it is useful to develop a general framework of scalar-tensor theories.\footnote{A complementary approach is the effective field theory (EFT) based on the spontaneous breaking of spacetime symmetries, which was originally developed in \cite{Creminelli:2006xe,Cheung:2007st} to describe the dynamics of cosmological perturbations in a model-independent manner. We note that the EFT of \cite{Creminelli:2006xe,Cheung:2007st} applies only to scalar-tensor theories, and it was extended to incorporate vector-tensor theories~\cite{Aoki:2021wew} and solids/fluids~\cite{Aoki:2022ipw}. Moreover, applications to black hole perturbations have also been discussed recently~\cite{Franciolini:2018uyq,Hui:2021cpm,Mukohyama:2022enj,Khoury:2022zor,Mukohyama:2022skk}.}
Another reason to study such generalization is that it gives rise to novel interactions in general, which are expected to exhibit peculiar phenomena that can be tested observationally.
The most general class of scalar-tensor theories with second-order Euler-Lagrange equations is known as the Horndeski class~\cite{Horndeski:1974wa,Deffayet:2011gz,Kobayashi:2011nu}.
Due to this second-order nature, the Horndeski class is obviously free from the problem of the Ostrogradsky ghost, i.e., unstable extra DOFs associated with higher-order equations of motion (EOMs)~\cite{Woodard:2015zca}.
However, the Horndeski class is not the most general class of ghost-free scalar-tensor theories:
One can go beyond the Horndeski class by use of disformal transformations~\cite{Bekenstein:1992pj,Bruneton:2007si,Bettoni:2013diz}.
A disformal transformation is a redefinition of the metric containing the scalar field and its derivatives, which includes the conformal transformation as a special case.
If one performs such a metric redefinition on some given scalar-tensor theories, then one obtains possibly new theories.
Up to the first-order derivative of the scalar field, one can write down the most general possible transformation law for the metric as
    \be
    \bar{g}_\mn[g_{\alpha\beta},\phi] = F_0(\phi,X) g_\mn + F_1(\phi,X)\na_\mu\phi\na_\nu\phi,
    \label{disformal1_intro}
    \ee
which maps a pair~$(g_{\alpha\beta},\phi)$ to $(\bar g_{\alpha\beta},\phi)$ with $\bar{g}_\mn$ being a new metric.
Here, $F_0$ and $F_1$ are functions of $(\phi,X)$ with $X\coloneqq g^{\alpha\beta}\pa_\alpha\phi\pa_\beta\phi$.
This transformation maps a theory within the Horndeski class to its outside in general~\cite{BenAchour:2016fzp}, which we shall call {\it the disformal Horndeski class} (or the DH class for short).
An important thing to note is that the transformation~\eqref{disformal1_intro} is invertible (i.e., can be solved uniquely for $g_{\alpha\beta}$ at least locally in the configuration space) so long as $F_0(F_0+XF_1)(F_0-XF_{0X}-X^2F_{1X})\ne 0$ is satisfied~\cite{Zumalacarregui:2013pma,Takahashi:2021ttd}, where a subscript~$X$ denotes the partial derivative with respect to $X$.
When such an invertible transformation acts on scalar-tensor theories, it does not change the number of physical DOFs, and in particular, it maps a ghost-free theory to another ghost-free theory~\cite{Domenech:2015tca,Takahashi:2017zgr}.
Therefore, thanks to the ghost-free nature of the Horndeski class, the DH class is also ghost-free.
Note that there are other classes of ghost-free scalar-tensor theories~\cite{Langlois:2015cwa,Crisostomi:2016czh,BenAchour:2016fzp,Takahashi:2017pje,Langlois:2018jdg} that are not disformally related to the Horndeski class.
These classes are free from the Ostrogradsky ghost since they satisfy the so-called degeneracy condition~\cite{Motohashi:2014opa,Langlois:2015cwa,Motohashi:2016ftl,Klein:2016aiq,Kimura:2017gcy,Motohashi:2017eya,Motohashi:2018pxg,Kimura:2018sfs} in a more nontrivial way.
However, these classes do not accommodate a viable cosmological solution~\cite{Langlois:2017mxy,Takahashi:2017pje,Langlois:2018jdg}, and hence they are phenomenologically disfavored.
This is why we are interested in ghost-free theories generated by disformal transformations.
Recently, a higher-derivative generalization of disformal transformations was developed in \cite{Takahashi:2021ttd} and was employed to obtain a novel class of ghost-free scalar-tensor theories, which is dubbed as {\it the generalized disformal Horndeski class} (or the GDH class)~\cite{Takahashi:2022mew}.
The GDH class provides the most general framework for ghost-free scalar-tensor theories to this date.

By definition, a (G)DH theory is related to some Horndeski theory via invertible (generalized) disformal transformation.
On the other hand, when minimally coupled matter fields are introduced to each of the two gravitational theories, the two systems are no longer disformal relatives.
In this sense, the existence of matter fields is crucial to distinguish the (G)DH class from the Horndeski class.
Also, gravity can only be probed through matter fields, and hence one should take them into account in practice.
We note that matter coupling in scalar-tensor theories (or any other modified gravity theories) is in general a nontrivial issue in the following sense:
Even though the gravitational sector satisfies the degeneracy condition and is free from the Ostrogradsky ghost, the introduction of matter fields can revive the ghost since the degeneracy structure is altered in general. 
We would like to avoid this problem at least for the standard model particles, i.e., a scalar field, gauge fields, and spinor fields with canonical kinetic terms.
Fortunately, such ordinary matter fields can be consistently coupled to DH theories without reviving the Ostrogradsky ghost~\cite{Deffayet:2020ypa}.
Further, the consistency of the matter coupling in the GDH class was partially addressed.
The authors of \cite{Takahashi:2022mew,Naruko:2022vuh} showed that ordinary bosonic matter fields can be consistently coupled under the unitary gauge in a particular subclass of the GDH class.
Namely, in this subclass, minimally coupled bosonic matter fields do not spoil the degeneracy structure (see \S\ref{ssec:GDT_metric} and \S\ref{ssec:bosonic} for details).
Still, the consistency of fermionic matter coupling in the GDH class has been remained a nontrivial issue so far.

In the present paper, we  
investigate the consistency of matter coupling in GDH theories, with a particular focus on fermionic matter coupling.
The most nontrivial thing in this case is that the fermion Lagrangian is written in terms of not the metric itself but the tetrad, whose transformation law under the generalized disformal transformation has not been clarified in the literature.
In this paper, we develop the transformation law for the tetrad to study the structure of the fermion kinetic term.
Interestingly, we find that the consistency of the fermionic matter coupling requires an additional condition on top of the one required for the bosonic case.
As a result, we find that the GDH class accommodates a subclass where not only bosonic but also fermionic matter fields can be consistently coupled without reviving the Ostrogradsky ghost.

The rest of this paper is organized as follows.
In \S\ref{sec:GDT}, we review the generalized disformal transformation and develop the tetrad transformation law associated with the generalized disformal transformation.
In \S\ref{sec:coupling}, we discuss the consistency of matter coupling in GDH theories based on the transformation law for the metric/tetrad obtained in \S\ref{sec:GDT}, where we reveal the additional condition necessary for making the fermionic as well as bosonic matter coupling consistent.
Finally, we draw our conclusions in \S\ref{sec:conc}.

\section{Generalized disformal transformation}\label{sec:GDT}

In this section, we briefly review the higher-derivative generalization of disformal transformations based on \cite{Takahashi:2021ttd,Takahashi:2022mew}.
We then develop the transformation law for the tetrad associated with the generalized disformal transformation.
We shall use the results of this section to discuss the consistency of matter coupling in \S\ref{sec:coupling}.
Note that our main focus is on a four-dimensional spacetime, but the results in this section hold in arbitrary spacetime dimensions.
If one is interested in higher dimensions, the term~``tetrad'' should be interpreted as ``vielbein.''

\subsection{Metric transformation law}\label{ssec:GDT_metric}

Let us consider the following generalized disformal transformation~\cite{Takahashi:2021ttd}:
    \be
    \bar{g}_\mn[g_{\alpha\beta},\phi]=F_0g_\mn+F_1\phi_\mu\phi_\nu+2F_2\phi_{(\mu}X_{\nu)}+F_3X_\mu X_\nu, \label{GDT}
    \ee
with the coefficients~$F_i$ being functions of $(\phi,X,Y,Z)$ and
    \be
    X\coloneqq \phi_\mu\phi^\mu, \qquad
    Y\coloneqq \phi_\mu X^\mu, \qquad
    Z\coloneqq X_\mu X^\mu.
    \ee
Here and in what follows, we use the notations~$\phi_\mu\coloneqq \pa_\mu\phi$ and $X_\mu\coloneqq \pa_\mu X$.
Note that the right-hand side of \eqref{GDT} is the most general symmetric tensor of rank two constructed out of $\phi$, $\phi_\mu$, and $X_\mu$.
Thanks to this structure, it is straightforward to construct the inverse metric~$\bar{g}^\mn$ such that $\bar{g}^{\mu\alpha}\bar{g}_{\alpha\nu}=\delta^\mu_\nu$~\cite{Takahashi:2021ttd}.
Moreover, as detailed in \cite{Takahashi:2021ttd}, one can systematically obtain the conditions under which the transformation is invertible [i.e., Eq.~\eqref{GDT} can be uniquely solved for the unbarred metric at least locally in the configuration space] by focusing on the group structure of the transformations under the functional composition.
In particular, the closedness under the functional composition requires that the coefficient functions~$F_i$ satisfy
    \be
    \bar{X}_Y=\bar{X}_Z=0, \qquad
    \bar{X}\coloneqq \bar{g}^\mn\phi_\mu\phi_\nu.
    \ee
Namely, the kinetic term~$\bar X$ of the scalar field associated with the barred metric, which is generically a function of $(\phi,X,Y,Z)$, must be a function only of $(\phi,X$):
    \be
    \bar{X}=\bar{X}(\phi,X). \label{Xbar(phi,X)}
    \ee
With this requirement, one can construct the inverse transformation in a fully covariant manner.
One can also see that, in order for the inverse transformation to exist, the following conditions should be satisfied on top of \eqref{Xbar(phi,X)}:
    \be
    F_0\ne 0, \qquad
    \mF\ne 0, \qquad
    \bar{X}_X\ne 0, \qquad
    \left|\fr{\pa(\bar{Y},\bar{Z})}{\pa(Y,Z)}\right|\ne 0,
    \label{inv_cond}
    \ee
where we have defined $\bar{Y}\coloneqq \bar{g}^\mn\phi_\mu\bar{X}_\nu$, $\bar{Z}\coloneqq \bar{g}^\mn\bar{X}_\mu\bar{X}_\nu$ with $\bar{X}_\mu\coloneqq\pa_\mu\bar{X}$, and
    \be
    \mF\coloneqq F_0^2+F_0(XF_1+2YF_2+ZF_3)+(XZ-Y^2)(F_1F_3-F_2^2). \label{mF}
    \ee

In \cite{Takahashi:2022mew}, the generalized disformal transformation~\eqref{GDT} satisfying the invertibility conditions~\eqref{Xbar(phi,X)} and \eqref{inv_cond} was employed to construct a novel class of ghost-free scalar-tensor theories, which is dubbed as {\it the generalized disformal Horndeski class} (or the GDH class for short).
A nontrivial thing in GDH theories (or any other ghost-free theories with degenerate higher-derivative interactions) is that the Ostrogradsky ghost could revive in general when matter fields are taken into account.\footnote{We expect that the mass of the revived Ostrogradsky ghost would scale as some inverse power of the matter energy density~$\rho$. Therefore, from the EFT point of view, the ghost would be irrelevant at low energies for sufficiently small $\rho$.}
Interestingly, the authors of \cite{Takahashi:2022mew,Naruko:2022vuh} showed that a subclass of GDH theories generated by generalized disformal transformations of the following form allows for consistent bosonic matter coupling:
    \be
    \bar{g}_\mn[g_{\alpha\beta},\phi]=f_0g_\mn+f_1\phi_\mu\phi_\nu+2f_2\phi_{(\mu}\DX{\nu)}+f_3\DX{\mu}\DX{\nu}, \qquad
    \DX{\mu}\coloneqq \bra{\delta_\mu^\alpha-\fr{\phi_\mu\phi^\alpha}{X}}\pa_\alpha X,
    \label{GDT_consistent}
    \ee
where $\DX{\mu}$ denotes the projection of $\pa_\mu X$ onto a constant-$\phi$ hypersurface and $f_i$'s are functions of $(\phi,X,\mZ)$ with\footnote{The quantity~$W\coloneqq Y^2-XZ$ used in \cite{Takahashi:2022mew} plays essentially the same role as $\mZ$ in the present paper.}
    \be
    \mZ\coloneqq \DX{\mu}\DXup{\mu}=Z-\fr{Y^2}{X}.
    \ee
Note that $f_i$'s here are related to $F_i$'s in \eqref{GDT} as
    \be
    F_0=f_0, \qquad
    F_1=f_1-\fr{2Y}{X}f_2+\fr{Y^2}{X^2}f_3, \qquad
    F_2=f_2-\fr{Y}{X}f_3, \qquad
    F_3=f_3.
    \label{f2F}
    \ee
We shall discuss the issue of matter coupling in detail in \S\ref{sec:coupling}.

In what follows, we employ the strategy of \cite{Takahashi:2021ttd} to derive several formulae for the inverse metric and the inverse transformation associated with the generalized disformal metric~\eqref{GDT_consistent}, which are useful when we focus on this particular subclass of generalized disformal transformations.
The expression for the inverse disformal metric~$\bar{g}^\mn$ is obtained by assuming
    \be
    \bar{g}^\mn[g_{\alpha\beta},\phi]=a_0g^\mn+a_1\phi^\mu\phi^\nu+2a_2\phi^{(\mu}\DXup{\nu)}+a_3\DXup{\mu}\DXup{\nu},
    \label{inverse_metric_spatial}
    \ee
and then fixing each coefficient function~$a_i(\phi,X,\mZ)$ so that $\bar{g}^{\mu\alpha}\bar{g}_{\alpha\nu}=\delta^\mu_\nu$.
Written explicitly, we have
    \be
    a_0=\fr{1}{f_0}, \qquad
    a_1=-\fr{f_0f_1+\mZ(f_1f_3-f_2^2)}{f_0\mF}, \qquad
    a_2=-\fr{f_2}{\mF}, \qquad
    a_3=-\fr{f_0f_3+X(f_1f_3-f_2^2)}{f_0\mF},
    \ee
with $\mF$ defined in \eqref{mF}, which can be expressed in terms of $f_i$'s as
    \be
    \mF=f_0^2+f_0\bra{Xf_1+\mZ f_3}+X\mZ\bra{f_1f_3-f_2^2}.
    \ee
As a result, we obtain
    \be
    \bar{X}
    =Xa_0+X^2a_1
    =\fr{X\bra{f_0+\mZ f_3}}{\mF},
    \ee
which is a function of $(\phi,X,\mZ)$ in general.
In order for the disformal transformation to be invertible, we require that $\bar{X}$ does not depend on $\mZ$, i.e., $\bar{X}=\bar{X}(\phi,X)$.
This requirement poses a constraint among $f_i$'s, which allows us to express one of the $f_i$'s in terms of the others.
For instance, $f_1$ is written in terms of the other functions as
    \be
    f_1=\fr{1}{\bar{X}}-\fr{f_0}{X}+\fr{\mZ f_2^2}{f_0+\mZ f_3}.
    \ee
We also assume $\bar{X}_X\ne 0$ so that we have $X$ as a function of $(\phi,\bar{X})$.
Let us define
    \be
    \bDX{\mu}\coloneqq \bra{\delta_\mu^\alpha-\fr{\phi_\mu\phi_\nu}{\bar{X}}\bar{g}^{\nu\alpha}}\pa_\alpha \bar{X}
    =\bar{X}_X\bra{\DX{\mu}+\fr{\mZ f_2}{f_0+\mZ f_3}\phi_\mu},
    \ee
and then
    \be
    \bar{\mZ}\coloneqq \bar{g}^\mn\bDX{\mu}\bDX{\nu}
    =\fr{X\bar{X}_X^2}{\bar{X}}\fr{\mZ}{\mF}.
    \ee
Provided that $(\mZ/\mF)_\mZ\ne 0$, this equation allows us to express $\mZ$ as a function of $(\phi,\bar{X},\bar{\mZ})$.
Now, the inverse disformal transformation can be written in the form
    \be
    g_\mn[\bar{g}_{\alpha\beta},\phi]
    =\fr{1}{f_0}\bar{g}_\mn-\fr{1}{f_0}\brb{f_1-\fr{\mZ f_2^2(2f_0+\mZ f_3)}{(f_0+\mZ f_3)^2}}\phi_\mu\phi_\nu-\fr{2f_2}{\bar{X}_X(f_0+\mZ f_3)}\phi_{(\mu}\bDX{\nu)}-\fr{f_3}{\bar{X}_X^2f_0}\bDX{\mu}\bDX{\nu}, \label{inverse_GDT_spatial}
    \ee
where $X$ and $\mZ$ on the right-hand side should be respectively regarded as functions of $(\phi,\bar{X})$ and $(\phi,\bar{X},\bar{\mZ})$, as explained earlier.
The invertibility conditions are summarized as
    \be
    f_0\ne 0, \qquad
    \mF\ne 0, \qquad
    \bar{X}_X\ne 0, \qquad
    \bar{X}_\mZ=0, \qquad
    \bra{\fr{\mZ}{\mF}}_\mZ\ne 0. \label{inv_cond_spatial}
    \ee
Note that Eqs.~\eqref{inverse_metric_spatial}--\eqref{inv_cond_spatial} are consistent with those in \cite{Takahashi:2021ttd} under the identification~\eqref{f2F}.

\subsection{Tetrad transformation law}\label{ssec:tetrad}

As mentioned earlier, we shall discuss the consistency of matter coupling in the presence of fermionic matter fields, whose action is written in terms of the tetrad~$e^a_\mu$.
Therefore, in this subsection, we develop the transformation law for the tetrad~$\bar{e}^a_\mu=\bar{e}^a_\mu[e^c_\alpha,\phi]$ so that $\bar{g}_\mn=\eta_{ab}\bar{e}^a_\mu\bar{e}^b_\nu$, with $\bar{g}_\mn$ being the generalized disformal metric studied in \S\ref{ssec:GDT_metric}.
Note that, instead of starting from the general form~\eqref{GDT}, we assume $\bar{g}_\mn$ to be of the form~\eqref{GDT_consistent} from the outset, since otherwise bosonic matter coupling revives the Ostrogradsky ghost~\cite{Takahashi:2022mew,Naruko:2022vuh}.
For this purpose, let us put the barred tetrad in the following form:\footnote{One can verify that \eqref{barred-tetrad_spatial} can be uniquely solved for the unbarred tetrad at least locally in the configuration space if the disformal transformation of the metric is invertible (see the \hyperref[AppA]{Appendix}).}
    \be
    \bar{e}^a_\mu[e^c_\alpha,\phi]
    =\bra{E_0\delta^\alpha_\mu+E_1\phi_\mu\phi^\alpha+E_2\phi_\mu \DXup{\alpha}+E_3\phi^\alpha\DX{\mu}+E_4\DX{\mu}\DXup{\alpha}}e^a_\alpha,
    \label{barred-tetrad_spatial}
    \ee
with the coefficients~$E_I$ ($I=0,1,\cdots,4$) being functions of $(\phi,X,\mZ)$.
Then, we have
    \begin{align}
    \bar{g}_\mn
    &=\eta_{ab}\bar{e}^a_\mu\bar{e}^b_\nu \nonumber \\
    &=E_0^2g_\mn
    +\bra{XE_1^2+\mZ E_2^2+2E_0E_1}\phi_\mu\phi_\nu
    +2\brb{E_0(E_2+E_3)+XE_1E_3+\mZ E_2E_4}\phi_{(\mu}\DX{\nu)} \nonumber \\
    &\quad +\bra{XE_3^2+\mZ E_4^2+2E_0E_4}\DX{\mu}\DX{\nu}.
    \end{align}
By equating this expression with \eqref{GDT_consistent}, we have
    \be
    \begin{split}
    &E_0^2=f_0, \qquad
    XE_1^2+\mZ E_2^2+2E_0E_1=f_1, \\
    &E_0(E_2+E_3)+XE_1E_3+\mZ E_2E_4=f_2, \qquad
    XE_3^2+\mZ E_4^2+2E_0E_4=f_3,
    \end{split}
    \label{tetrad_coeffs}
    \ee
where the first equation requires $f_0>0$.
Note that we have only four equations for five unknown variables~$E_I$, meaning that the system is underdetermined.
Nevertheless, the ambiguity due to the underdeterminedness should be absorbed into a local Lorentz transformation and hence does not affect the action for a fermionic matter field, as we shall discuss below.
By making use of this ambiguity, one can make $E_3=0$.
We then obtain
    \be
    E_0=\sqrt{f_0}, \qquad
    E_1=\fr{\sqrt{X/\bar{X}}-\sqrt{f_0}}{X}, \qquad
    E_2=\fr{f_2}{\sqrt{f_0+\mZ f_3}}, \qquad
    E_3=0, \qquad
    E_4=\fr{\sqrt{f_0+\mZ f_3}-\sqrt{f_0}}{\mZ}.
    \label{coeffs_noNdot}
    \ee
Here, we assumed $E_0>0$ and chose a branch such that $\bar{e}^a_\mu=E_0 e^a_\alpha$ (i.e., $E_1=E_2=E_4=0$) for the case of conformal transformations, where $f_1=f_2=f_3=0$ and $\bar{X}=X/f_0$.
Now, the transformation law~\eqref{barred-tetrad_spatial} for the tetrad is written only in terms of $\phi$, $\phi_\mu$, and $\DX{\mu}$, and hence the time derivative of the lapse function is absent under the unitary gauge.
Note also that, as they should be, the barred tetrad satisfies $\bar{g}^\mn\bar{e}^a_\mu\bar{e}^b_\nu=\eta^{ab}$ and its dual~$\bar{e}^\mu_a$ (i.e., such that $\bar{g}^\mn=\eta^{ab}\bar{e}^\mu_a\bar{e}^\nu_b$) is given by $\bar{e}^\mu_a=\bar{g}^\mn\eta_{ab}\bar{e}^b_\nu$.

Let us now study the local Lorentz transformation.
We assume that the transformation has the form
    \be
    \bar{e}^a_\mu \to \Lambda^a{}_b\bar{e}^b_\mu, \qquad
    \Lambda^a{}_b\coloneqq e^a_\mu \bra{L_0\delta^\mu_\nu+L_1\phi^\mu\phi_\nu+L_2\phi^\mu\DX{\nu}+L_3\phi_\nu\DXup{\mu}+L_4\DXup{\mu}\DX{\nu}}e^\nu_b,
    \ee
with $L_I$'s being functions of $(\phi,X,\mZ)$, such that
    \be
    \eta_{cd}\Lambda^c{}_a\Lambda^d{}_b=\eta_{ab}.
    \ee
This equation poses the following constraints on $L_I$'s:
    \be
    \begin{split}
    &L_0^2=1, \qquad
    XL_1^2+\mZ L_3^2+2L_1=0, \qquad
    L_2+L_3+XL_1L_2+\mZ L_3L_4=0, \qquad
    XL_2^2+\mZ L_4^2+2L_4=0.
    \end{split} \label{local_Lorentz}
    \ee
From the first equation, we can choose $L_0=1$.
Then, the latter three equations should be satisfied by the remaining four functions, meaning that there exists a family of local Lorentz transformations characterized by one function of $(\phi,X,\mZ)$.
Acting the local Lorentz transformation~$\Lambda^a{}_b$ on the barred tetrad in \eqref{barred-tetrad_spatial}, we have
    \begin{align}
    \Lambda^a{}_b\bar{e}^b_\mu
    =\bra{E_0\delta^\alpha_\mu+\ti{E}_1\phi_\mu\phi^\alpha+\ti{E}_2\phi_\mu\DXup{\alpha}+\ti{E}_3\phi^\alpha\DX{\mu}+\ti{E}_4\DX{\mu}\DXup{\alpha}}e^a_\alpha,
    \end{align}
where
    \be
    \begin{split}
    &\ti{E}_1=(1+XL_1)E_1+L_1E_0+\mZ L_2E_2, \qquad
    \ti{E}_2=\bra{1+\mZ L_4}E_2+L_3E_0+XL_3E_1, \\
    &\ti{E}_3=(1+XL_1)E_3+L_2E_0+\mZ L_2E_4, \qquad
    \ti{E}_4=\bra{1+\mZ L_4}E_4+L_4E_0+XL_3E_3.
    \end{split}
    \ee
As it should be, one can confirm that the left-hand sides of \eqref{tetrad_coeffs} are invariant under the local Lorentz transformation.
Hence, one can make $\ti{E}_3=0$ by choosing $L_1$ such that
    \be
    (1+XL_1)E_3+L_2E_0+\mZ L_2E_4=0. \label{local_Lorentz_E3=0}
    \ee
Combining \eqref{local_Lorentz} and \eqref{local_Lorentz_E3=0}, we have the following solution for $L_I$'s:
    \be
    \begin{split}
    XL_1=\mZ L_4&=-1+\brb{1+X\mZ\bra{\fr{E_3}{E_0+\mZ E_4}}^2}^{-1/2}, \\
    L_2=-L_3&=-\fr{E_3}{E_0+\mZ E_4}\brb{1+X\mZ\bra{\fr{E_3}{E_0+\mZ E_4}}^2}^{-1/2}.
    \end{split}
    \ee
Note that we chose a branch such that the local Lorentz transformation approaches identity (i.e., $L_1=L_2=L_3=L_4=0$) in the limit~$E_3\to 0$.
This guarantees that one can impose $E_3=0$ without loss of generality in obtaining \eqref{coeffs_noNdot}.

The above discussion clarifies that the tetrad transformation law associated with the generalized disformal transformation~\eqref{GDT_consistent} is given by \eqref{barred-tetrad_spatial} with \eqref{coeffs_noNdot}.
A caveat is that there still remain ambiguities, even at the starting point~\eqref{barred-tetrad_spatial}:
One may note that one could incorporate yet higher derivatives, e.g.,
    \be
    \mZ_\mu\coloneqq \bra{\delta^\alpha_\mu-\fr{\phi_\mu\phi^\alpha}{X}}\pa_\alpha\mZ,
    \ee
into the tetrad transformation law~\eqref{barred-tetrad_spatial}, while keeping the metric transformation law~\eqref{GDT_consistent}.
We expect that such terms would also be absorbed into a (generalized) local Lorentz transformation.
Since it is more convenient for our purpose to use a simpler form of tetrad transformations, we adopt \eqref{barred-tetrad_spatial} with \eqref{coeffs_noNdot} for the following arguments rather than studying more involved forms.

\section{Consistency of matter coupling}\label{sec:coupling}

Let us now consider matter fields which are minimally coupled to GDH theories.
By construction of the GDH class, one can move to the frame where the gravitational action reduces to the Horndeski one while the matter fields are coupled to the generalized disformal metric, i.e.,
    \be
    S[g_\mn,\phi,\Psi]
    \coloneqq S_{\rm H}[g_\mn,\phi]+S_{\rm m}[\bar{g}_\mn,\Psi], \label{Horndeski_frame}
    \ee
with the matter fields being denoted collectively by $\Psi$.
Note that the (barred/unbarred) metric here should be replaced by the (barred/unbarred) tetrad if the matter sector contains fermionic components.
As shown in \cite{Takahashi:2022mew}, this ``Horndeski frame'' is useful to study the consistency of matter coupling.
We note that the invertibility of generalized disformal transformations plays a crucial role in going back and forth between the two frames.
We revisit the case of bosonic matter fields in \S\ref{ssec:bosonic} and discuss the case of fermionic matter fields in \S\ref{ssec:fermionic}.\footnote{One could consider matter fields that are coupled to Horndeski gravity through the generalized disformal metric corresponding to noninvertible transformations [more precisely, those that do not satisfy at least one of the last three conditions in Eq.~\eqref{inv_cond_spatial}].
Our discussion in \S\ref{ssec:bosonic} and \S\ref{ssec:fermionic} also applies to such a case, except that one cannot move to the frame where the matter fields are minimally coupled to GDH theories.
Interestingly, without the invertibility conditions, one could obtain a weaker set of conditions for consistent matter coupling~\cite{Naruko:2022vuh}.}

\subsection{Bosonic matter}\label{ssec:bosonic}

Let us first study the case of bosonic matter fields coupled to GDH theories.
We assume that the scalar field~$\phi$ in the gravitational sector has a timelike gradient and adopt the unitary gauge where $\phi=\phi(t)$.
Also, we introduce the Arnowitt-Deser-Misner (ADM) variables as
    \be
    g_\mn \D x^\mu \D x^\nu=-N^2\D t^2+h_{ij}(\D x^i+N^i\D t)(\D x^j+N^j\D t), \label{ADM}
    \ee
where $N$ is the lapse function, $N^i$ is the shift vector, and $h_{ij}$ is the induced metric.

It is important to note that the Lagrangian of ordinary bosonic matter fields does not contain derivatives of the metric to which they are minimally coupled.
For instance, the action of a canonical scalar field~$\sigma$ is given by
    \be
    S_{\rm scalar}[g_\mn,\sigma]=\int \D^4x\sqrt{-g}\brb{-\fr{1}{2}g^\mn\pa_\mu\sigma\pa_\nu\sigma-V(\sigma)}, \label{S_scalar}
    \ee
where no derivative acts on $g_\mn$.
The same is true for the following action of a canonical gauge field~$A^A_\mu$:
    \be
    S_{\rm gauge}[g_\mn,A^A_\mu]=\int \D^4x\sqrt{-g}\bra{-\fr{1}{4}F^A_\mn F^{A\mn}}, \qquad
    F^A_\mn\coloneqq \pa_\mu A^A_\nu-\pa_\nu A^A_\mu+g_{*}f^{ABC}A^B_\mu A^C_\nu, \label{S_gauge}
    \ee
where $g_{*}$ denotes the gauge coupling constant and $f^{ABC}$ is the structure constant associated with the gauge group.
One could of course consider a matter field whose Lagrangian contains derivatives of the metric (e.g., cubic Galileon as studied in \cite{Deffayet:2020ypa,Iyonaga:2021yfv}), but we are mostly interested in ordinary matter fields like those described by the actions~\eqref{S_scalar} and \eqref{S_gauge} or at most their nonlinear generalizations (e.g., k-essence scalar field~\cite{ArmendarizPicon:1999rj} and nonlinear electrodynamics~\cite{Born:1934gh,Heisenberg:1936nmg}).

Now, let us consider matter fields that are coupled to the generalized disformal metric, as we mentioned above in \eqref{Horndeski_frame}.
As clarified in \cite{Takahashi:2022mew,Naruko:2022vuh}, under the unitary gauge, the generalized disformal metric~\eqref{GDT} in general contains the time derivative of the lapse function~$N$ and so does the matter action~$S_{\rm m}[\bar{g}_\mn,\Psi]$, which makes $N$ dynamical when coupled to Horndeski gravity.
This is the origin of the revival of the Ostrogradsky ghost in the presence of matter fields.
On the other hand, restricted to the subclass of generalized disformal transformations given by \eqref{GDT_consistent}, the matter coupling is consistent, i.e., the Ostrogradsky ghost does not revive.
To see this, recall that the transformation~\eqref{GDT_consistent} is constructed out of $\phi$, $\phi_\mu$, and $\DX{\mu}$.
Under the unitary gauge, the quantity~$\DX{\mu}$ corresponds to the three-dimensional covariant derivative of $X=-\dot{\phi^2}/N^2$ on a constant-$t$ hypersurface.
Therefore, the matter action does not yield the kinetic term of $N$, and hence the matter coupling is consistent.
In the language of the original frame, GDH theories that are generated via the generalized disformal transformation of the form~\eqref{GDT_consistent} allow for consistent bosonic matter coupling.\footnote{The condition for consistent matter coupling becomes tighter in scalar-tensor theories with a nondynamical scalar field (e.g., the cuscuton~\cite{Afshordi:2006ad} or its extension~\cite{Iyonaga:2018vnu,Iyonaga:2020bmm}), as pointed out in \cite{Takahashi:2022mew}.}

\subsection{Fermionic matter}\label{ssec:fermionic}

Having discussed the consistency of bosonic matter coupling, let us consider a fermionic matter field.
For concreteness, we consider a free massless Dirac spinor~$\psi$ in a curved spacetime described by the following action~\cite{Freedman:2012zz}:\footnote{Note that the discussion here would apply to the case of massive Dirac fields as well as Majorana fields.}
    \be
    S_{\rm spinor}[e^a_\mu,\psi]=\int \D^4x\,e\bra{-\fr{1}{2}\psi^\dagger {\rm i}\ga^{\hat{0}}e^\mu_a\ga^a\na_\mu\psi+{\rm c.c.}}, \label{fermion}
    \ee
where $e\coloneqq \det e^a_\mu$, ${\rm c.c.}$ denotes the complex conjugate, and $\gamma^a$ denotes the gamma matrices in the Minkowski spacetime such that $\gamma^a\gamma^b+\gamma^b\gamma^a=2\eta^{ab}\mathbbm{1}$, with $\mathbbm{1}$ being the identity matrix in the spinor indices.
Note that we put hats on local Lorentz indices ($a,b,\cdots=\{\hat{0},\hat{1},\hat{2},\hat{3}\}$).
Here, we do not use the notation $\bar{\psi}$.  Usually, it denotes $\psi^\dagger {\rm i}\gamma^{\hat{0}}$ in the literature, but in the present paper we reserve $\bar{\psi}$ to denote a transformation of the spinor field [see Eq.~\eqref{psibar}].
The covariant derivative acting on the Dirac field is defined by
    \be
    \na_\mu\psi
    \coloneqq \bra{\mathbbm{1}\pa_\mu+\fr{1}{4}\omega_{\mu}{}^{ab}\ga_{ab}}\psi.
    \ee
Here, $\ga_{ab}\coloneqq (\ga_a\ga_b-\ga_b\ga_a)/2$ and the (torsion-free) spin connection~$\omega_{\mu}{}^{ab}$ is defined by
    \be
    \omega_\mu{}^a{}_b=-e_b^\nu\bra{\pa_\mu e^a_\nu-\Gamma^\lambda_\mn e^a_\lambda},
    \ee
where $\Gamma^\lambda_\mn$ is the Christoffel symbol associated with the metric.
Equivalently, in terms of differential forms, the spin connection 1-form~$\omega^{ab}\coloneqq \omega_\mu{}^{ab}\D x^\mu$ is defined so that
    \be
    \D e^a=-\omega^a{}_b\wedge e^b, \label{spin_connection}
    \ee
with $e^a\coloneqq e^a_\mu\D x^\mu$ being the tetrad 1-form.

Since $\omega_\mu{}^{ab}$ involves derivatives of the tetrad/metric, one may think that derivatives of the lapse function and/or shift vector show up in the matter Lagrangian, which could make the fermionic matter coupling inconsistent even in Horndeski theories.
Nevertheless, this is not the case as we explain below.
For the ADM metric~\eqref{ADM}, the tetrad 1-form can be explicitly written as
    \be
    e^{\hat{0}}=N\D t, \qquad
    e^{\hat{i}}=\tri^{\hat{i}}_k(\D x^k+N^k\D t),
    \label{tetrad_ADM}
    \ee
where $\tri^{\hat{i}}_k$ denotes the triad such that $h_{kl}=\delta_{\hat{i}\hat{j}}\tri^{\hat{i}}_k\tri^{\hat{j}}_l$.
From \eqref{tetrad_ADM}, we have
    \be
    \D e^{\hat{0}}=\pa_kN\,\D x^k\wedge \D t, \qquad
    \D e^{\hat{i}}=\brb{\tridot^{\hat{i}}_k-\pa_k\bra{\tri^{\hat{i}}_lN^l}}\D t\wedge \D x^k+\pa_k\tri^{\hat{i}}_l\,\D x^k\wedge \D x^l,
    \ee
where the time derivative (denoted by a dot) acts only on the triad associated with the induced metric.
Comparing these equations with \eqref{spin_connection}, it is clear that the time derivative of $N$ or $N^i$ does not appear in the spin connection.

On the other hand, the time derivative of the triad shows up through
    \be
    \omega_{\hat{i}\hat{0}}
    \supset \fr{1}{N}\tri^{\hat{j}}_k\tri_{({\hat{i}}}^l\tridot_{{\hat{j}})l}\bra{\D x^k+N^k\D t}, \qquad
    \omega_{\hat{i}\hat{j}}
    \supset \fr{1}{N}\tri_{[{\hat{i}}}^k\tridot_{{\hat{j}}]k}\D t.
    \ee
This means that, the spinor action would yield the time derivative of $N$ after the disformal transformation.
Indeed, from \eqref{barred-tetrad_spatial}, the transformation law for the triad is obtained as
    \be
    \btri^{\hat{i}}_k=\bra{E_0\delta^l_k+E_4\DX{k}\DXup{l}}\tri^{\hat{i}}_l, \qquad
    \btri_{\hat{i}}^k=\brb{\fr{1}{E_0}\delta^k_l-\fr{E_4}{E_0(E_0+\mZ E_4)}\DXup{k}\DX{l}}\tri_{\hat{i}}^l,
    \label{barred-triad}
    \ee
where the right-hand sides involve the lapse function (or its spatial derivatives through $\DX{k}$ and $\mZ$).
As a result, the spin connection associated with the barred tetrad yields $\dot{N}$, which could make the spinorial matter coupling inconsistent.

In order to investigate the degeneracy structure in detail, let us focus on terms in the spinor action~\eqref{fermion} that involve the time derivative of the spinor and the triad:
    \be
    S_{\rm spinor}[e^a_\mu,\psi]
    \supset \int\D^4x\sqrt{h}\bra{\fr{\rm i}{2}\psi^\dagger\dot{\psi}-\fr{\rm i}{2}\dot{\psi}^\dagger\psi+\fr{\rm i}{4}\psi^\dagger\tri_{\hat{i}}^k\tridot_{\hat{j}k}\gamma^{\hat{i}\hat{j}}\psi},
    \label{spinor_kinetic}
    \ee
with $h\coloneqq \det h_{kl}=(\det \tri^{\hat{i}}_k)^2$.
It should be noted that the right-hand side does not depend on $N$ or $N^i$.
Substituting the barred triad into the above equation, we obtain
    \begin{align}
    S_{\rm spinor}[\bar{e}^a_\mu,\psi]
    \supset \int\D^4x\sqrt{h}\,E_0^2(E_0+\mZ E_4)\bigg[&\fr{\rm i}{2}\psi^\dagger\dot{\psi}-\fr{\rm i}{2}\dot{\psi}^\dagger\psi+\fr{\rm i}{4}\psi^\dagger\tri_{\hat{i}}^k\tridot_{\hat{j}k}\gamma^{\hat{i}\hat{j}}\psi \nonumber \\
    &-{\rm i}\fr{X^2\mZ E_4^2}{E_0(E_0+\mZ E_4)}\fr{\pa_kN\pa_l\dot{N}}{N^2}\psi^\dagger\tri_{\hat{i}}^k\tri_{\hat{j}}^l\gamma^{\hat{i}\hat{j}}\psi\bigg], \label{spinor_kinetic_bar}
    \end{align}
where we have employed \eqref{barred-triad} and the following formula:
    \be
    \sqrt{\bar{h}}=\sqrt{h}\,E_0^2(E_0+\mZ E_4).
    \ee
One can absorb the overall factor~$E_0^2(E_0+\mZ E_4)$ by redefining the spinor~$\psi$ as
    \be \label{psibar}
    \psi=E_0^{-1}(E_0+\mZ E_4)^{-1/2}\bar{\psi},
    \ee
where, as mentioned above, we use $\bar{\psi}$ to denote the new spinor field rather than $\psi^\dagger {\rm i}\gamma^{\hat{0}}$.
Then, \eqref{spinor_kinetic_bar} takes the form
    \begin{align}
    S_{\rm spinor}[\bar{e}^a_\mu,\psi]
    \supset \int\D^4x\sqrt{h}\bigg[&\fr{\rm i}{2}\bar{\psi}^\dagger\dot{\bar{\psi}}-\fr{\rm i}{2}\dot{\bar{\psi}}^\dagger\bar{\psi}+\fr{\rm i}{4}\bar{\psi}^\dagger\tri_{\hat{i}}^k\tridot_{\hat{j}k}\gamma^{\hat{i}\hat{j}}\bar{\psi}
    -{\rm i}\fr{X^2\mZ E_4^2}{E_0(E_0+\mZ E_4)}\fr{\pa_kN\pa_l\dot{N}}{N^2}\bar{\psi}^\dagger\tri_{\hat{i}}^k\tri_{\hat{j}}^l\gamma^{\hat{i}\hat{j}}\bar{\psi}\bigg].
    \label{spinor_kinetic_bar2}
    \end{align}
Note that, except for the last term, the right-hand side has the same form as \eqref{spinor_kinetic}.

Let us study how $\dot{\bar{\psi}}$ and $\dot{N}$ show up in the EOMs.
The variation of the action with respect to $\bar{\psi}^\dagger$ involves the following terms:
    \be
    \fr{-{\rm i}}{\sqrt{\gamma}}\fr{\delta S}{\delta \bar{\psi}^\dagger}
    \supset \dot{\bar{\psi}}-\fr{X^2\mZ E_4^2}{E_0(E_0+\mZ E_4)}\fr{\pa_kN\pa_l\dot{N}}{N^2}\tri_{\hat{i}}^k\tri_{\hat{j}}^l\gamma^{\hat{i}\hat{j}}\bar{\psi}.
    \label{psi_EOM}
    \ee
Hence, the EOM for $\bar{\psi}^\dagger$, together with its Hermitian conjugate, yields\footnote{We choose a representation of the gamma matrices such that $\gamma^{\hat{i}}$'s are Hermitian matrices.}
    \be
    \label{dpgp} \pa_t\bra{\bar{\psi}^\dagger\gamma^{\hat{i}\hat{j}}\bar{\psi}}=\text{(terms without $\dot{\bar{\psi}}$ or $\dot{N}$)}.
    \ee
Likewise, the EOM for $N$ contains $\dot{\bar{\psi}}$, $\dot{\bar{\psi}}^\dagger$, and $\dot{N}$ in general. 
To avoid an unwanted extra DOF, these equations should be degenerate, i.e., there should exist an appropriate linear combination of the EOMs for $N$ and $\bar{\psi}$ that gives a constraint equation where any of $\dot{\bar{\psi}}$, $\dot{\bar{\psi}}^\dagger$, or $\dot{N}$ does not appear.
However, we see that this is not the case:
By using \eqref{dpgp}, one can erase $\dot{\bar{\psi}}$ and $\dot{\bar{\psi}}^\dagger$ simultaneously from the EOM for $N$, but $\dot{N}$ still remains, which cannot be removed by use of \eqref{psi_EOM} without reintroducing $\dot{\bar{\psi}}$.
This problem originates from the last term in \eqref{spinor_kinetic_bar2}, which is proportional to $E_4^2$.
This means that there exists an unwanted extra DOF so long as $E_4\ne 0$, i.e., $f_3\ne 0$ [see Eq.~\eqref{coeffs_noNdot}].
Conversely, if $f_3=0$, then \eqref{spinor_kinetic_bar2} has the same form as \eqref{spinor_kinetic}, and hence the spinor can be coupled to gravity without reviving the Ostrogradsky ghost.

To summarize, the consistency of the spinorial matter coupling requires an additional condition~$f_3=0$ compared to the bosonic case.
Thus, we identified a class of GDH theories to which both bosonic and fermionic matter fields can be consistently coupled.
Written explicitly, the generalized disformal metric after imposing the consistency of matter coupling takes the form
    \be
    \bar{g}_\mn[g_{\alpha\beta},\phi]=f_0g_\mn+f_1\phi_\mu\phi_\nu+2f_2\phi_{(\mu}\DX{\nu)},
    \label{GDT_consistent_fermion}
    \ee
where $f_i$'s are functions of $(\phi,X,\mZ)$.
Here, in order for the transformation to be invertible, the function~$f_1$ should be related to $f_0$ and $f_2$ as
    \be
    f_1=\fr{1}{\bar{X}}-\fr{f_0}{X}+\fr{\mZ f_2^2}{f_0},
    \ee
through some function~$\bar{X}=\bar{X}(\phi,X)$.
We note that the class of generalized disformal transformations of the form~\eqref{GDT_consistent_fermion} covers conventional disformal transformations (i.e., those with at most first-order derivative of $\phi$).

\section{Conclusions}\label{sec:conc}

There have been extensive studies on scalar-tensor theories with higher-order derivatives, where the degeneracy of higher-derivative terms is crucial for avoiding the problem of the Ostrogradsky ghost.
The largest class of ghost-free scalar-tensor theories known so far was proposed recently in \cite{Takahashi:2022mew} and is called the generalized disformal Horndeski (GDH) class, which is generated from the class of Horndeski theories via invertible generalized disformal transformation~\cite{Takahashi:2021ttd}.
When matter fields are coupled to such gravitational theories, the Ostrogradsky ghost can revive as the matter Lagrangian does not necessarily respect the degeneracy conditions imposed on the gravitational Lagrangian.
Needless to say, matter fields exist in our Universe, and hence one should construct gravitational theories so that the matter fields can be consistently coupled without reviving the Ostrogradsky ghost.
It was shown in \cite{Takahashi:2022mew,Naruko:2022vuh} that ordinary bosonic matter fields (e.g., a k-essence scalar field and standard gauge fields) can be consistently coupled to a particular subclass of GDH theories.
This subclass is obtained via the generalized disformal transformation~\eqref{GDT_consistent}, which does not involve the time derivative of the lapse function in the unitary gauge (see \S\ref{ssec:GDT_metric} and \S\ref{ssec:bosonic} for a detailed discussion).

On the other hand, in the case of fermionic matter fields, the matter Lagrangian is written in terms of not the metric itself but the tetrad, and hence the analyses in \cite{Takahashi:2022mew,Naruko:2022vuh} do not apply directly.
Therefore, in the present paper, we studied the tetrad transformation law under the generalized disformal transformation in \S\ref{ssec:tetrad}.
We showed that, if the generalized disformal transformation does not depend on the time derivative of the lapse function under the unitary gauge, neither does the associated tetrad transformation law.
By use of the transformation law for the tetrad, we studied how the action of a spinorial matter field is transformed under the generalized disformal transformation.
We clarified that an additional condition [i.e., $f_3=0$ in the notation of \eqref{GDT_consistent}] is required in order not to revive the Ostrogradsky ghost (see \S\ref{ssec:fermionic} for details).

To summarize, we identified a class of GDH theories that allows for consistent coupling of both bosonic and fermionic matter fields.
Although the consistency of matter coupling puts a tight constraint on the form of the generalized disformal transformation and GDH theories associated with it, there still remains a nontrivial class of GDH theories that includes the DH class as a strict subset.
Our strategy would be straightforwardly applied to more generic invertible transformations as well as other modified gravity theories, e.g., vector-tensor theories and multi-field scalar-tensor theories.
It should also be noted that our analysis (as well as those in \cite{Takahashi:2022mew,Naruko:2022vuh}) relies on the unitary gauge where the scalar field has a spatially uniform profile.
Indeed, the consistency of matter coupling under the unitary gauge would be enough for the GDH theory to be employed in the cosmological context where the scalar field typically has a timelike profile.
Away from the unitary gauge, there would be a shadowy mode~\cite{DeFelice:2018mkq,DeFelice:2021hps,DeFelice:2022xvq} in matter-coupled GDH theories in general, which itself is harmless.
Having said that, the analysis away from the unitary gauge would also be important when one is interested in a situation where the scalar profile is spacelike, which we expect to happen for, e.g., neutron star solutions.\footnote{Black holes/neutron stars in shift-symmetric scalar-tensor theories can support scalar hair with a timelike profile~\cite{Mukohyama:2005rw,Babichev:2013cya,Kobayashi:2014eva,Babichev:2016jom,Sakstein:2016oel,Tretyakova:2017lyg,Chagoya:2018lmv,Kobayashi:2018xvr,Motohashi:2019sen,Charmousis:2019vnf,Takahashi:2019oxz,Takahashi:2020hso,Takahashi:2021bml,Ikeda:2021skk,Mukohyama:2022enj,Mukohyama:2022skk,DeFelice:2022qaz}, for which one can take the unitary gauge.}
If one is interested in such a situation, one has to study the structure of the kinetic matrix without assuming the unitary gauge, which should pose a tighter constraint on the transformation law.
We hope to address this issue in a future work~\cite{Ikeda:2023ntu}.


\acknowledgments{
This work was supported by JSPS (Japan Society for the Promotion of Science) KAKENHI Grant No.~JP21J00695 (K.T.), No.~JP22K03605 (R.K.), and No.~JP22K03639 (H.M.).
}


\appendix*

\section{Inverse transformation law for the tetrad}\label{AppA}

In this Appendix, we briefly discuss the inverse transformation associated with the tetrad transformation~\eqref{barred-tetrad_spatial}.
On top of \eqref{barred-tetrad_spatial}, let us consider a disformal tetrad given by
    \be
    \ti{e}^a_\mu[e^c_\alpha,\phi]
    =\bra{A_0\delta^\alpha_\mu+A_1\phi_\mu\phi^\alpha+A_2\phi_\mu \DXup{\alpha}+A_3\phi^\alpha\DX{\mu}+A_4\DX{\mu}\DXup{\alpha}}e^a_\alpha, \label{tilded-tetrad}
    \ee
which will be identified as the inverse transformation for \eqref{barred-tetrad_spatial}.
The functional composition of the two disformal transformations yield
    \begin{align}
    \ti{e}^a_\mu[\bar{e}^c_\alpha,\phi]
    &=\bra{\ti{A}_0\delta^\alpha_\mu+\ti{A}_1\phi_\mu\phi^\alpha+\ti{A}_2\phi_\mu \DXup{\alpha}+\ti{A}_3\phi^\alpha\DX{\mu}+\ti{A}_4\DX{\mu}\DXup{\alpha}} \nonumber \\
    &\quad\times\bra{E_0\delta^\beta_\alpha+E_1\phi_\alpha\phi^\beta+E_2\phi_\alpha \DXup{\beta}+E_3\phi^\beta\DX{\alpha}+E_4\DX{\alpha}\DXup{\beta}}e^a_\beta, \label{tetrad_composition}
    \end{align}
where we have reorganized terms inside the first parentheses in terms of unbarred quantities.
One can confirm that the map between $A_I$'s in \eqref{tilded-tetrad} and $\ti{A}_I$'s in \eqref{tetrad_composition} is invertible if the disformal transformation for the metric is invertible.
The inverse transformation for \eqref{barred-tetrad_spatial} is obtained by putting $\ti{e}^a_\mu[\bar{e}^c_\alpha,\phi]=e^a_\mu$, which yields a system of linear algebraic equations for $\ti{A}_I$'s.
There exists a unique solution to this system if and only if the following quantity does not vanish:
    \be
    E_0^2+E_0(XE_1+\mZ E_4)+X\mZ(E_1E_4-E_2E_3)\ne 0. \label{inv_cond_tetrad}
    \ee
By use of \eqref{tetrad_coeffs}, one can show that
    \be
    \mF=\brb{E_0^2+E_0(XE_1+\mZ E_4)+X\mZ(E_1E_4-E_2E_3)}^2,
    \ee
which is nonvanishing under the condition~\eqref{inv_cond_spatial}.
This means that the condition~\eqref{inv_cond_tetrad} is automatically satisfied if the generalized disformal transformation for the metric is invertible.
Therefore, the invertibility of the tetrad transformation follows from that of the metric transformation.


\bibliographystyle{mybibstyle}
\bibliography{bib}

\end{document}